\documentclass[letterpaper,twocolumn,prl,showpacs]{revtex4-1}
\usepackage{amsmath,amssymb,bbold}
\usepackage{graphicx}
\usepackage{bm,braket}
\usepackage{xcolor}

\newcommand{\br}{\boldsymbol{r}}
\newcommand{\bulk}{\scriptscriptstyle{\text{(bulk)}}}
\newcommand{\surf}{\scriptscriptstyle{\text{(surf)}}}
\newcommand{\site}{\scriptscriptstyle{\text{site}}}
\newcommand{\bbm}{\mathbb{m}}
\newcommand{\bbc}{\mathbb{c}}
\newcommand{\avgb}[1]{\bigl\langle\, #1\, \bigr\rangle_{{}_{\!\!\scriptscriptstyle{\text{bulk}}}}}
\renewcommand{\[}{\begin{equation}}
\renewcommand{\]}{\end{equation}}
\def\bea{\begin{eqnarray}}
\def\eea{\end{eqnarray}}
\def\nn{\nonumber\\}

\newcommand{\emi}[1]{{\rm e}^{-i #1}}
\newcommand{\ei}[1]{{\rm e}^{i #1}}
\newcommand{\B}{{\bf B}}
\newcommand{\dk}{[d \, {\rm k}]}
\newcommand{\intk}{\int_{\rm BZ} \!\!\!\! [d \, {\rm k}] \;}

\newcommand{\M}{{\bf M}}
\newcommand{\Mb}{{\bf M}^{\scriptscriptstyle{\text{(bulk)}}}}
\newcommand{\m}{{\bf m}}
\newcommand{\CQ}{{\cal Q}}
\newcommand{\CP}{{\cal P}}

\newcommand{\q}{{\bf k}}
\renewcommand{\k}{{\bf k}}
\newcommand{\vc}{V_{\rm cell}}
\renewcommand{\r}{{\bf r}}

\newcommand{\da}{\partial_\alpha}
\newcommand{\db}{\partial_\beta}

\newcommand{\equ}[1]{Eq.~(\ref{#1})}
\newcommand{\eqs}[2]{Eqs.~(\ref{#1}) and (\ref{#2})}
\newcommand{\s}{\mbox{\boldmath$\tilde{\sigma}$}}

\def\bra#1{\langle#1\vert}
\def\ket#1{\vert#1\rangle}
\def\ev#1{\langle#1\rangle}
\def\me#1#2#3{\langle#1| \, #2 \, |#3\rangle}
\def\runtime{(\the\time)\qquad\the\month/\the\day/\the\year}% get current time
\def\today
 {\count10=\year\advance\count10 by -2000 \number\day--\ifcase
  \month \or Jan\or Feb\or Mar\or Apr\or May\or Jun\or
             Jul\or Aug\or Sep\or Oct\or Nov\or Dec\fi--\number\count10}

\def\hour{\count10=\time\count11=\count10
\divide\count10 by 60 \count12=\count10
\multiply\count12 by 60 \advance\count11 by -\count12\count12=0
\number\count10 :\ifnum\count11 < 10 \number\count12\fi\number\count11}

\begin{document}

%%%%%%%%%%%%%%%%%%%%%%%%%%%%%%%%%%%%%%%%%
\title{Orbital Magnetization in Insulators: Bulk vs. Surface}

\author{Raffaello Bianco$^1$ and Raffaele Resta$^{2,3}$ \\ ~~}

\affiliation{\centerline{$^1$ CNRS, UMR 7590, and
Sorbonne Universit\'es, UPMC Univ Paris 06, 4 place Jussieu, F-75005, Paris, France}}
\affiliation{\centerline{$^2$ Dipartimento di Fisica, Universit\`a di Trieste, 34127 Trieste, Italy}}
\affiliation{\centerline{$^3$ Donostia International Physics Center, 20018 San Sebasti\'an, Spain}}

\date{\today}

\begin{abstract}
The orbital magnetic moment of a finite piece of matter is expressed in terms of the one-body density matrix as a simple trace. We address a macroscopic system, insulating in the bulk, and we show that its orbital moment is the sum of a bulk term and a surface term, both extensive. The latter only occurs when the transverse conductivity is nonzero and owes to conducting surface states. Simulations on a model Hamiltonian validate our theory.
\end{abstract}

\date{run through \LaTeX\ on \today\ at \hour}

\pacs{xxx}

\maketitle \bigskip\bigskip

According to magnetostatics, the orbital magnetic moment is defined as $\m = -\partial E/\partial \B$. 
This applies to any {\it bounded} piece of matter. For a homogeneous macroscopic
system of volume $V$ one writes $\m = V \M$, where $\M$ is the  macroscopic magnetization. 
While in trivial insulators only the bulk states contributes to the magnetic moment,
in nontrivial ones (defined below) a term coming
from the surface states appears:
\begin{equation}
\m=\m^{\bulk}+\m^{\surf},
\label{def}
\end{equation}
and dividing by $V$ we obtain the corresponding contributions to the
orbital magnetization $\M^{\bulk}$ and $\M^{\surf}$.

In this paper we propone a new approach to study the orbital
magnetization, based on an observable which allows to discriminate
the separate contributions to the total magnetic moment coming from the
surface and bulk states. It applies to any kind 
of insulator, crystalline or noncrystalline.
As we will see, a key property of this approach is that it
is free from the drawbacks related to the use of currents. 

We define as ``nontrivial'' any insulator having a non-vanishing transverse conductivity at zero $\B$ field, which we encode in the vector quantity $\tilde{\sigma}_\gamma = \epsilon_{\gamma\alpha\beta}\sigma_{\alpha\beta}/2$, where $\varepsilon_{\gamma\alpha\beta}$ is the antisymmetric tensor, and the sum over Cartesian indices is implicit. We stress that we are primarily addressing {\it noncrystalline}---although macroscopically homogeneous---systems; it is nonetheless straightforward to extend our treatment to inhomogeneous systems where $\M^{\bulk}$ varies in space over a macroscopic length scale. 
In any homogeneous insulator the $T=0$ longitudinal conductivity vanishes, while in general $\s \neq 0$. For a 2D system $\s$ is a dimensionless integer when expressed in $e^2/h$ units, while for a 3D system it has the dimension of an inverse length, and is quantized only in the crystalline case. 

The difficulties in defining what
$\M$ {\it really is}, and what is the role played by the
edge states, are closely related to the use of currents. 
Edge states arise due to the existence of a confining potential
but disappear as we consider periodic
boundary conditions (PBCs). In the thermodynamic
limit they do not contribute to the density of states per unit volume,
but the orbital magnetization can be affected by them.
However, even in systems where only the bulk electrons contribute
to the orbital magnetization, the currents which appear at the boundary
must be taken into account in order to estimate the correct value for the
magnetic moment. This consideration has also been at the root of the longstanding debate 
about the bulk nature of the orbital magnetization.The problem is  
well emphasized in the classic review by Hirst \cite{Hirst97}:
a {\it finite} magnetized sample is characterized by a dissipationless current
flowing at its boundary, but in an {\it unbounded} sample (as addressed in condenser
matter physics) the macroscopic orbital magnetization $\M$ is apparently indeterminate.
In that paper, Hirst analyzes the problem in terms of microscopic current densities
(either classical or quantum), and summarizes the state of the art at the time of publication (1997). 

It is clear nowadays that the quantum Hamiltonian (and the corresponding ground state) are explicitly
needed in order to define and to compute $\M$ for an unbounded sample: the bulk microscopic current density
is not enough. In fact, it has been shown in 2005-06 \cite{Xiao05,rap128,rap130} that for a crystalline
sample within PBCs:  
\bea  M_\gamma &=& - \frac{ie}{2 \hbar c} \epsilon_{\gamma\alpha\beta} \times \label{chern} \\ &  \times &\sum_{\varepsilon_{j\k}< \mu} \intk \me{\da u_{j\q}}{ (H_\k + \epsilon_{j\q} - 2 \mu )}{\db u_{j\q}}.  \nonumber \eea  In \equ{chern} BZ is the Brillouin zone, $\mu$ is the Fermi level, Greek subscripts are Cartesian indices, $\da = \partial/\partial k_\alpha$,  $\dk = d \k /(2\pi)^d$ where d is the dimensionality (either 2 0r 3), $\ket{u_{j\q}} = \emi{\k \cdot \r} \ket{\psi_{j\q}}$ are the lattice-periodic factors in the Bloch orbitals, normalized over the unit cell of volume (area in 2D) $\vc$; they are eigenfunctions of  $H_{\q} = \emi{\q \cdot \r} H \ei{\q \cdot \r}$, with eigenvalues $\epsilon_{j\k}$. 

The existence of a formula for the orbital magnetization
within PBCs clarifies that, in principle, the orbital magnetization can 
be obtained by considering only the bulk of a material.
However, the relation between this formula and the standard definition
of the magnetic moment within ``open'' boundary conditions (OBCs), 
is still somewhat obscure. Moreover, the different roles played by
edge and bulk states are not clearly characterized.
In fact, the explicit dependence of $\M$ from the value of $\mu$ in the bulk,
in Eq.~\eqref{chern}, implies the presence of edges and, possibly, the contact with an \textit{external}
electron \textit{reservoir} in order to change $\mu$ as a control parameter. 
It has been observed that this is rather puzzling, since Eq.~\eqref{chern}
addresses a system with no edges~\cite{Chen11}, whereas any experiment (even \textit{gedanken})
addresses a bounded sample.
The approach presented here 
sheds light on these apparently paradoxical aspects:
we provide an alternative expression for $\M$
and, from the very beginning, we get rid of currents, which are not the good observable
in order to analyze the different role played by bulk and surface states.
In fact, surface currents are due to both bulk and surface states and, in general,
it is not trivial to disentangle the two contributions.
As we will see, in a sense a simple analog of our transformation is an integration by part: 
the same integral can be obtained from very different integrands.

The key observation for the following of the present work is that \equ{chern} applies as it stands even to a {\it noncrystalline} system,
provided we adopt a very large supercell, and consequently a mini-BZ. In fact we are adopting here the same supercell viewpoint upon
which the topological nature of the quantum Hall effect was established \cite{Thouless82, Kohmoto85}.
While in any crystalline insulator the spectrum $\epsilon_{j\k}$ is gapped, the spectral gap might close in the large supercell limit.
This happens for an Anderson insulator, where the supercell size must be ideally larger than the Anderson localization length. 
We observe that \equ{chern} retains its validity even for gapless materials \cite{rap130}, and that the case where the mini-BZ 
collapses to a single point has been studied in detail \cite{rap135}.

We address a macroscopically homogeneous, although possibly disordered, piece of matter. For any bounded independent-electron system within OBCs, the moment is, by definition: \[ \m = - \frac{\partial}{\partial \B} \sum_{j=1}^N \epsilon_j = - \frac{e}{2c} \sum_{j=1}^N \me{\varphi_j}{\r \times {\bf v}}{\varphi_j} \label{total} .\] 
We neglect any spin-dependent property here, and we deal with ``spinless electrons''; ${\bf v}$ is the quantum-mechanical velocity operator, $\epsilon_j$, $\ket{\varphi_j}$ are the single-particle eigenvalues and orbitals, and  $N$ is the number of electrons. \equ{total} is the circulation of the whole microscopic current density: bulk and surface.
 
We recast \equ{total} in trace form. To this aim we define the density matrix (a.k.a. ground state projector) $\CP$; we will also need the complementary projector $\CQ= {\cal I} - \CP$. Their definitions are \[ \CP =  \sum_{j=1}^N \ket{\varphi_j} \bra{\varphi_j} , \quad \CQ = \sum_{j=N+1}^\infty \ket{\varphi_j} \bra{\varphi_j}  .\] The $\gamma$ component of $\m$, from \equ{total}, is 
\[  m_\gamma =  - \frac{e}{2  c} \varepsilon_{\gamma\alpha\beta} \mbox{Tr} \{ \CP r_\alpha  v_\beta \} = - \frac{i e}{2 \hbar c} \varepsilon_{\gamma\alpha\beta} \mbox{Tr} \{ \CP r_\alpha  H  r_\beta \}, \label{tr1} \] where we have used ${\bf v} = i [H,\r]/\hbar$. Lengthy although straightforward manipulations of the trace in \equ{tr1} lead---exploiting antisymmetry--- to \cite{Souza08,rap148}: \[ m_\gamma = - \frac{i e}{2 \hbar c} \varepsilon_{\gamma\alpha\beta} \mbox{Tr} \{ \CP r_\alpha \CQ H \CQ r_\beta \CP  -  \CQ r_\alpha \CP  H  \CP r_\beta \CQ \} . \label{tr2} \] We then notice that---when the two traces are expressed in the Schr\"odinger representation---the integrated value over the whole sample is the same, but there is a key difference in the {\it integrands}. The {\it unbounded} position operator $\r$ is the essential ingredient of \equ{tr1}, which is therefore well defined only within OBCs, i.e. if the $\ket{\varphi_j}$ orbitals are square-integrable. Instead, only the projected $\r$ operator $\CP \r \CQ$ and its Hermitian conjugate $\CQ \r\CP$ enter \equ{tr2}: this has far reaching consequences. It is known since long time that such projected operators are well defined and regular even in an unbounded system within PBCs. Furthermore $\me{\r}{\CP \r \CQ}{\r'}$ is {\it nearsighted} \cite{Kohn96} in insulators, i.e. it decays exponentially (times a polynomial) for $|\r - \r'| \rightarrow \infty$. An important theorem proves the exponential decay even for Anderson insulators  \cite{Aizenman98}. 

Motivated by these considerations, we address the \textit{local marker} in real space
for the magnetic moment
\begin{equation}
\bbm_\gamma(\br)\equiv
- \frac{i e}{2 \hbar c} \varepsilon_{\gamma\alpha\beta} \Braket{\br|
\CP r_\alpha \CQ H \CQ r_\beta \CP  -  \CQ r_\alpha \CP  H  \CP r_\beta \CQ 
|\br} ,
\label{eq:local_marker}
\end{equation}
whose integral over the sample gives the total magnetic moment $m_\gamma$.
Thanks to the properties of the operator  $\CP\br\CQ$, the 
marker $\bbm_\gamma(\br)$ is well-defined with either OBCs or PBCs.
Moreover, for an insulator it is local in the bulk, since its value in a point $\br$
of the bulk is affected only by the electronic distribution on a region exponentially localized around it.

We obtain the bulk contribution to the orbital magnetization
simply by considering the average value of the marker (i.e. its integral per unit volume) 
in the bulk of the sample
\begin{equation}
M_{\gamma}^{\bulk}=\avgb{\bbm_\gamma},
\label{eq:def_bulk}
\end{equation}
because the value of $\bbm_\gamma(\br)$ in the bulk is independent of the boundary conditions adopted
and within PBCs we discard any surface effect.
Therefore, the surface contribution to the total magnetic moment, $\m^{\surf}=$$\m-\m^{\bulk}$, 
is given by the local marker $\bbm_\gamma(\br)$ on the surface of the sample (when OBCs are adopted).
As a consequence, the value of $\bbm(\br)$ on the surface must be extensive, 
although the boundary region is not such. This counterintuitive feature has also been confirmed by
simulations at variable sample size (not presented here).
While the term  $\m^{\bulk}$ is the one ideally measurable by accessing the electron distribution in the
bulk of the sample only, $\m^{\surf}$ owes to the electron distribution in the boundary region
of the sample. As we will see later, $ \m^{\surf}$ is different from zero only for nontrivial insulators.

So far, we have implicitly considered an {\it isolated} bounded sample at fixed $N$, with the only requirement that the resulting Fermi level $\mu$ falls in a mobility gap;
next we are going to consider the same system in contact with an electron reservoir which controls the $\mu$ (and $N$) value.
We observe that the value of $\M^{\bulk}$ (and $\M^{\surf}$) clearly depends on the (arbitrary) energy zero, while 
the total value of $\M$ does not depend on it. It is therefore expedient to set the energy zero at the lowest bulk-gap edge: with this choice
the longitudinal conductivity is nonzero for negative $\mu$ and vanishes for positive $\mu$ (insofar as $\mu$ remains in the mobility gap). 

We have defined $\Mb$ as a quantity which can be computed (and ideally measured) in the bulk of a sample, either bounded or unbounded,  \equ{eq:def_bulk}; next we wish to retrieve $\Mb$ within the supercell approach of \equ{chern} in reciprocal space. We are going to show that \equ{eq:def_bulk}, coincides with \equ{chern}, when we set $\mu=0$ {\it in the integrand only}.  If localized states in the mobility gap are present this still has a $\mu$ dependence, as it must be. 
In order to prove the equivalence we need the explicit expression for $\CP \r \CQ$ and its Hermitian conjugate $\CQ \r\CP$ within PBCs. While the $\r$ operator itself is ill-defined, its off-diagonal elements are well defined: this is a staple of linear response in solids \cite{Baroni01}. One of the known expression is  \[ \me{\psi_{j' \q}}{\r}{\psi_{j \q'}} = i \frac{(2\pi)^{\rm d}}{\vc} \delta(\q - \q') \ev{u_{j'\q} | \nabla_{\q} u_{j\q}} , \quad j \neq j'\label{dipole} , \] where we stress that a supercell viewpoint is adopted here.
We may therefore express $\CQ \r \CP$ as \[ \CQ \r \CP = i  \vc \sum_{\varepsilon_{j\k}< \mu} \; \sum_{\varepsilon_{j'\k} > \mu} \intk \ket{\psi_{j'\q}}  \ev{u_{j'\q} | \nabla_{\q} u_{j\q}} \bra{ \psi_{j\q}} . \label{eq:PrQ_kspace} \]  The first of the operators entering the trace in \equ{tr2} becomes thus \bea && \CP r_\beta \CQ H \CQ r_\alpha \CP  = \vc \sum_{\varepsilon_{j\k}< \mu} \; \sum_{\varepsilon_{j'\k} > \mu} \intk \times \nn &\times& \ket{\psi_{j\q}}   \ev{\db u_{j\q} | u_{j'\q}} \epsilon_{j'\q} \ev{ u_{j'\q} | \da u_{j\q}}\bra{ \psi_{j\q}} .\eea  When taking the trace per cell, we may replace the sum over the conduction bands ($\varepsilon_{j'\k} > \mu$) with the sum over all bands, since the difference is a symmetric tensor. Exploiting completeness we arrive at
\bea  && \mbox{Tr} \{ \CP r_\beta \CQ H \CQ r_\alpha \CP \} \nn &=& \vc\sum_{\varepsilon_{j\k}< \mu} \intk \me{\db u_{j\q}}{ H_\k }{\da u_{j\q}} . \eea

Similar manipulations performed on the second term in the trace in \equ{tr2} lead to the final result \bea  M_\gamma^{\bulk} &=&  - \frac{ie}{2 \hbar c} \epsilon_{\gamma\alpha\beta} \times \label{normal} \\ &\times&  \sum_{\varepsilon_{j\k}< \mu} \intk \me{\da u_{j\q}}{ (H_\k + \epsilon_{j\q} )}{\db u_{j\q}}. \nonumber \eea 
By comparing \eqs{chern}{normal} to \equ{def}, we clearly get $\M^{\surf}$ by difference:
\begin{equation}
M_{\gamma}^{\surf} =\mu \frac{ie}{\hbar c} \epsilon_{\gamma\alpha\beta} \sum_{\varepsilon_{j\k}< \mu} \intk \Braket{\da u_{j\q}|\db u_{j\q}}.
\label{eq:msurf}
\end{equation}
For cristalline insulators, it can be written as
\begin{equation}
M_{\gamma}^{\surf} =\mu \frac{2\pi e}{\hbar c}C_\gamma ,
\label{eq:msurf1}
\end{equation}
where $C_\gamma$ is  the Chern invariant, defined as 
\begin{equation}
C_{\gamma}=\frac{i}{2\pi}\epsilon_{\gamma\alpha\beta} \sum_{\varepsilon_{j\k}< \mu} \intk \Braket{\da u_{j\q}|\db u_{j\q}} .
\end{equation}
However, for a generic insulator, by using~Eq.~\eqref{eq:PrQ_kspace} for $\CQ \br \CP$ (and its conjugate)
we can write Eq.~\eqref{eq:msurf} in real space \cite{rap146} as the average in the bulk of another 
local marker $\bbc_{\gamma}(\br)$
\begin{equation}
M_{\gamma}^{\surf} =\mu \frac{2\pi e}{\hbar c}\, \avgb{\bbc_{\gamma}},
\label{eq:msurf2}
\end{equation}
with
\begin{equation}
\bbc_{\gamma}(\br)=\frac{i}{4\pi} \varepsilon_{\gamma\alpha\beta}\Braket{\br|\CP r_\alpha \CQ  r_\beta \CP  -  \CQ r_\alpha\CP  r_\beta \CQ |\br} .
\label{eq:cmarker}
\end{equation}
Notice that Eq.~\eqref{eq:msurf2}, at variance with Eq.~\eqref{eq:msurf1}, is well defined both within OBCs and PBCs, 
and it has the same value regardless of the boundary conditions adopted. 
It allows to connect the surface term to transverse conductivity for a generic insulator.
Starting from the standard Kubo-Greenwood formula for the transverse conductivity of a system which in the bulk has a mobility gap,
straightforward manipulations allow to express even $\s$ in terms of $\CP \r \CQ$ and its Hermitian conjugate $\CQ \r\CP$ \cite{rap146},
and we obtain: \[ \s=-\frac{e c}{\mu}\,\,\M^{\surf}  . \label{msurf2} \]
Therefore, as anticipated, $ \M^{\surf}$ is different from zero for, and only for, nontrivial insulators,
that is for insulators having non-zero transverse conductivity. Instead, for trivial insulators $\M^{\bulk} = \M$.
While $\s$ is obviously a bulk property, $\M^{\surf}$ only depends on something which ``happens'' near its boundary. %, as we are going to show with our simulations. 
We stress the virtue of our approach: we are not dealing with boundary currents; we deal instead with the integrated moment due to the boundary states altogether.
\equ{msurf2} is an outstanding manifestation of bulk-boundary correspondence: in insulators with nonzero transverse conductivity, the bulk and the boundary are ``locked''. What appears as ``bulk'' within PBCs, becomes  indeed ``surface'' when addressing a bounded sample within OBCs.
For this reason the surface contribution to the total $\M$ can be ``smeared'' into the bulk of the sample, simply replacing $H$ with $H - \mu$ in 
the expression for the local marker, \equ{eq:local_marker}: thus even the term which is actually due to the surface states of the finite sample appears ``as if'' it were a bulk term. We thus recover the local formula for orbital magnetization proposed in Ref. \cite{rap148}.

Finally, we analyze the variation with $\mu$ in the gap of the quantities $\M^{\bulk}$ and $\M^{\surf}$.
Here we stress the key difference between a system having a spectral gap (such as a crystalline insulator)
and one having only a mobility gap (such as an Anderson insulator).
When the system is in contact with an electron reservoir, 
a $\mu$ variation cannot affect the bulk electron distribution in the former case, 
while the opposite occurs in the latter.
Therefore, we have a first result: as anticipated, $\M^{\bulk}$ is constant with respect to $\mu$
for crystalline insulators, whereas, in the most general case considered here, $\M^{\bulk}$ is
$\mu$-dependent via the projectors $\CP$ and $\CQ$, Eq.~\eqref{eq:local_marker}. 

In order to analyze the variation of $\M^{\surf}$ with $\mu$, we exploit the general relationship:
\begin{equation}
\s = -ec \,\frac{ \partial \M^{\surf} }{ \partial \mu} \; .
\label{eq:cond_msurf}
\end{equation}
For crystalline insulators~Eq.~\eqref{eq:cond_msurf} reduces to $\s = -ec \, \partial \M / \partial \mu$, while if there are localized states in the mobility gap
the equality no longer holds, and the extra term coming from $\partial \Mb / \partial \mu$ has to be discounted: as shown in Ref. \cite{Xiao06}
this term corresponds to a current {\it non measurable} in a transport experiment. 
By comparing Eq.~\eqref{eq:cond_msurf} with Eq.~\eqref{msurf2}, it is immediate to see that $\M^{\surf}$ depends linearly on $\mu$. This is a general result, valid for any kind of insulator (crystalline or not). 

\begin{figure}[b]
\centering
\includegraphics[width=\columnwidth]{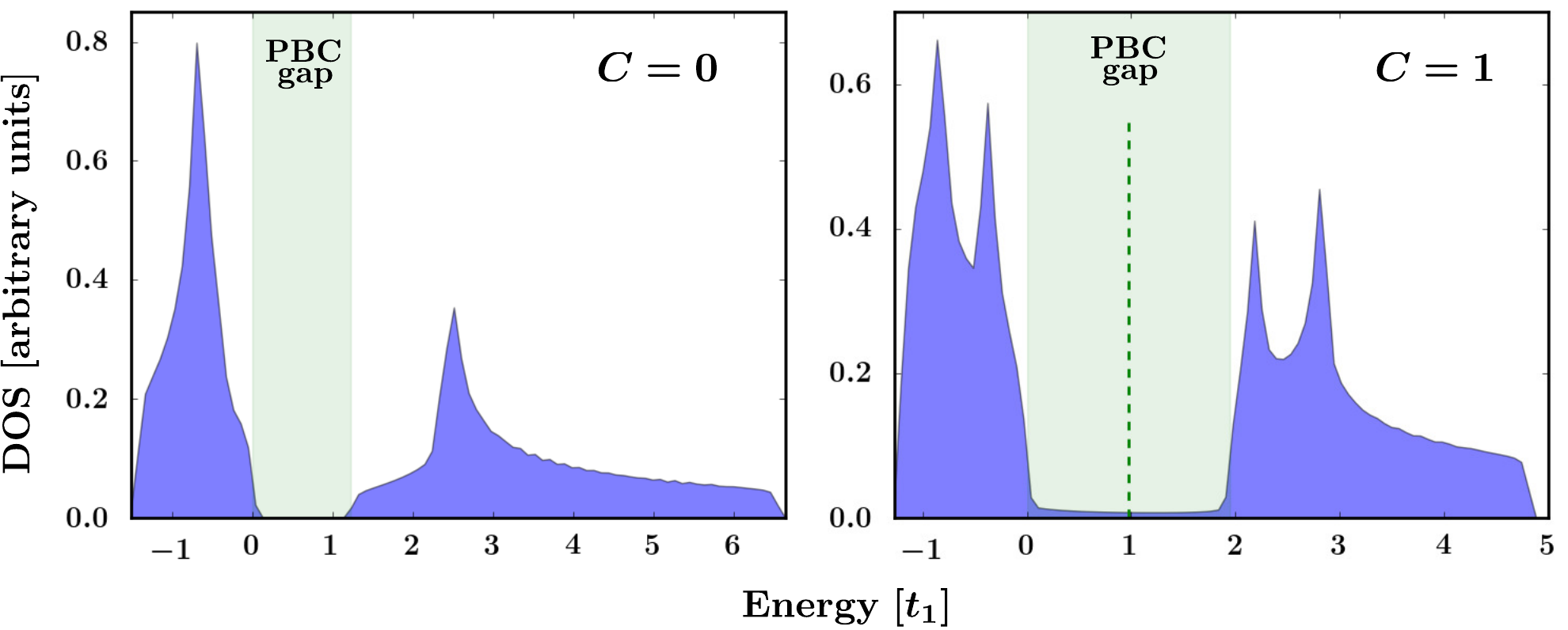}
\caption{(color online) Density of states for the Haldane model within OBCs (arbitrary units); the shaded area is the PBCs gap.  In both
cases $t_2=t_1/3$. 
{\it Left-hand panel}.  $\Delta=4\,t_2,\phi=0.1\,\pi$: trivial insulator with $\tilde{\sigma}_z=0$. 
{\it Right-hand panel.} $\Delta=t_2,\phi=0.4\,\pi$: nontrivial insulator with $\tilde{\sigma}_z = -e^2/h$. 
The nonzero value in the bulk gap owes to states at the boundary of the flake.  
The vertical dashed line marks the value $\mu=0.97\,t_1$ (see Fig. \ref{fig:additional}).
}
\label{fig:dos} \end{figure}

\begin{figure}[t]
\centering
\includegraphics[width=.85\columnwidth]{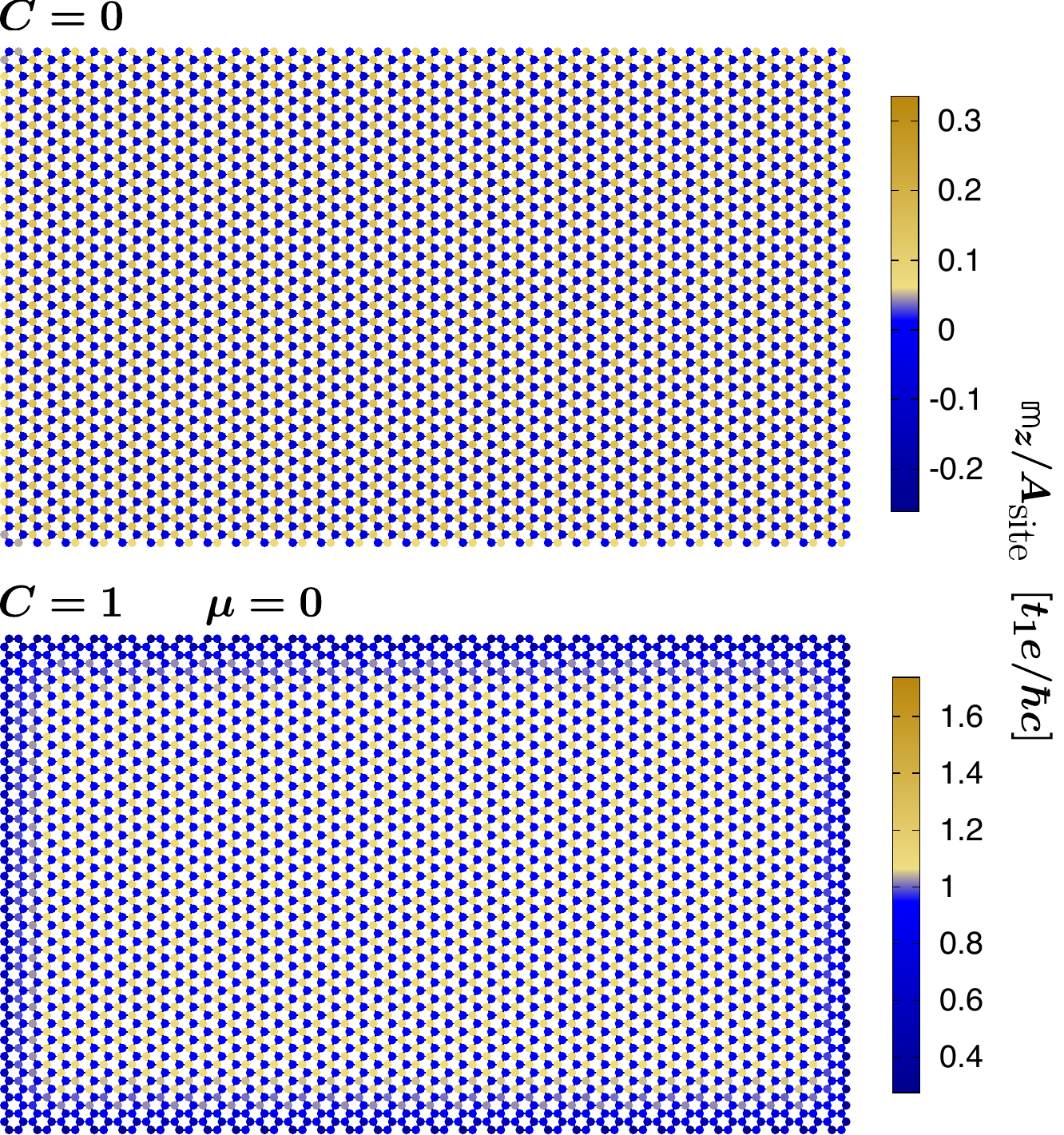}
\caption{(color online) Plot of $\bbm_z(\br)/A_{\site}$, where 
$\bbm_z(\br)$ is the local marker for the orbital magnetic moment along $z$, defined in
Eq.~\eqref{eq:local_marker}, and $A_{\site}$ is the sample area divided by the number of sites. 
The plot is in units  of $t_1 e /(\hbar c)$. {\it Top panel.} 
Trivial insulator, $\mu$ in the gap region. {\it Bottom panel.} Nontrivial insulator, $\mu=0$ at the lowest edge of the bulk gap.}
\label{fig:bulk} \end{figure}

Simulations addressing model Anderson insulators are notoriously very demanding \cite{Kramer93}. We can address here only a model system having a spectral gap: we therefore choose a 2D flake cut from a crystalline system. We adopt the Haldane Hamiltonian \cite{Haldane88}, also adopted by many papers including Refs. \cite{rap128,rap130,rap148,rap146}. The model is
comprised of a 2D honeycomb lattice with two tight-binding sites per primitive
cell with site energy difference of $2 \Delta$, real first-neighbor hoppings $t_1$, and
complex second-neighbor hoppings $t_2e^{\pm i\phi}$. According to the parameter values, the material may have $\tilde{\sigma}_z = -e^2/h, 0, e^2/h$. 
All of our simulations are performed for rectangular flakes with $N_{\site}$=3660 sites, within OBCs. 
We choose two representative cases; their density of states is shown in Fig. \ref{fig:dos} and their {\it bulk} magnetization, 
$M_z^{\bulk}=m^{\bulk}_z/A$, is $0.037$ and $1.006$ respectively, in units of $t_1 e /(\hbar c)$ (where $A$ is the flake area).

In a tight-binding model the total trace in \equ{tr2} becomes a discrete sum over the atomic sites of the 
local marker defined in Eq.~\eqref{eq:local_marker}. For the two samples, we show in Fig. \ref{fig:bulk} the local marker value for each site,
divided by $A_{\site}\equiv A/N_{\site}$. The average of this quantity returns the orbital magnetization $M_z$.
These site contributions are not gauge invariant, individually. Only the trace per unit area of the local marker is gauge-invariant
 and in principle measurable (see below): in this tight-binding model it obtains from the sum of any two nearest-neighbor contributions in the bulk region, divided by two. 
We have plotted Fig. \ref{fig:bulk} for both the trivial insulator
with an arbitrary $\mu$ value in the gap, and for the nontrivial one at $\mu=0$, i .e. at the bottom of the bulk gap. 
The absence of populated surface states manifests itself in a uniform value of the local marker value.
The presence of finite size effects---due to the discreteness of the spectrum---explains the 
very small spurious boundary contribution in the nontrivial case, magnified by the chosen energy scale.

\begin{figure}[t]
\centering
\includegraphics[width=.85\columnwidth]{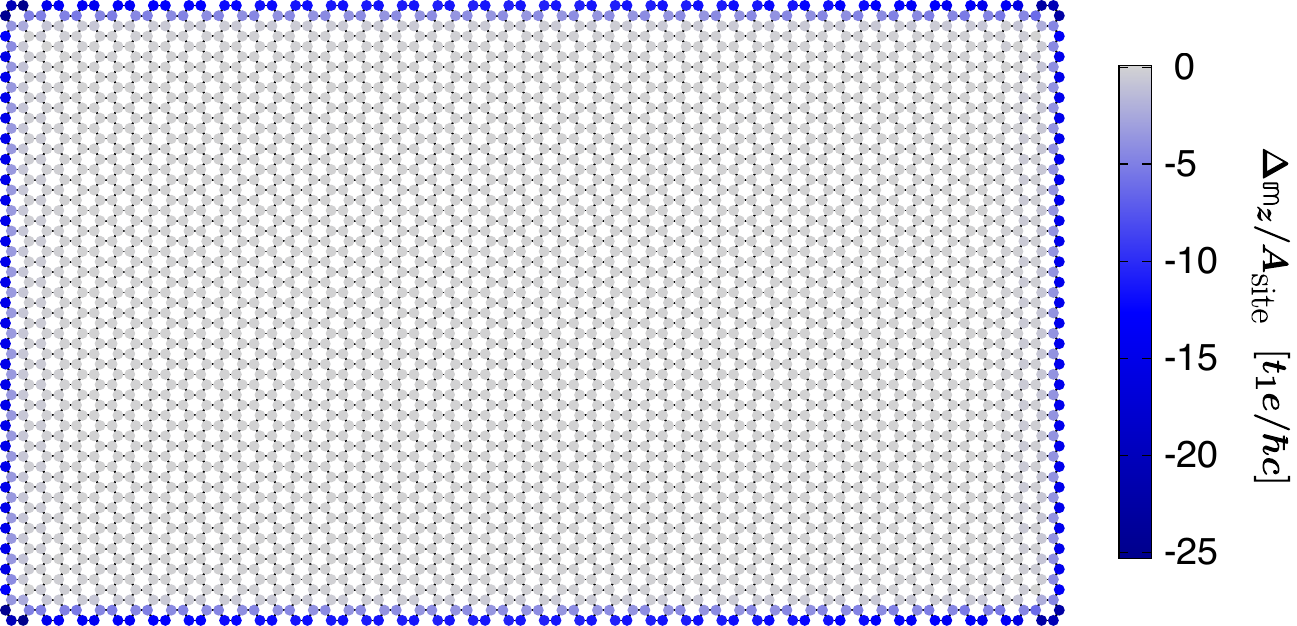}
\caption{(color online) Nontrivial insulator: plot of the difference $\Delta \bbm_z(\br)/A_{\site}$ 
between the cases $\mu=0.97\, t_1$ (dashed line in Fig.~\ref{fig:dos}) and $\mu=0$. 
Same units as in Fig.~\ref{fig:bulk}.}\label{fig:additional}
\end{figure}

Since we are considering a crystal, in the normal case a variation of $\mu$ in the gap cannot change anything. 
That is why, for the trivial insulator case, Fig.~\ref{fig:bulk} refers to an arbitrary $\mu$ value in the gap.
In the nontrivial case, instead, the filling of the boundary states provides a very large additional contribution $m_z^{\surf}$
which scales linearly with $\mu$, whereas the value in the bulk remains  unchanged. 
Our simulations confirm all these findings. The surface nature of the additional contribution is perspicuous 
in Fig.~\ref{fig:additional} (notice the different scales in Figs. \ref{fig:bulk} and \ref{fig:additional}). 

Our theory has addressed magnetization ``itself'' in macroscopically homogenous systems, either disordered or crystalline, 
while instead---as discussed in Ref. \cite{Hirst97}---one most often measures differences (or derivatives) in inhomogeneous situations.
Another virtue of the present approach is that it applies to inhomogeneous systems as well, where $\M$ varies on a macroscopic length scale.
The local marker in the real space is defined in the same way, 
through the diagonal of the relevant operators within Schr\"odinger representation: then 
the average per unit volume becomes the ``macroscopic average'' of that marker, defined as in electrostatics \cite{Jackson}.
In the special case of an heterojunction the system is insulating when $\mu$ is in the common bulk gap of the two materials.
Then \equ{eq:def_bulk} yields the bulk magnetizations, while \equ{eq:msurf2} accounts for a $\mu$-dependent interface term in nontrivial materials.

In conclusion, we show that the magnetic moment of a macroscopic piece of insulating matter is the sum two terms, both extensive,
due to states in the bulk and at the boundary of the system, respectively, and localized in the corresponding regions of the sample. 
The surface term only occurs when the the transverse conductivity is nonzero.
We stress that we are not addressing the current carried by these states: our approach---not based on currents---directly yields the moment due to bulk
and surface states, separately. 
The approach presented has a clear connection with the basic definition of magnetic moment for finite systems
and analyzes, with a common formalism, both crystalline and non crystalline insulators.
We have illustrated our theory with a simulation for a crystalline system; nonetheless, as said, the theory applies as well to an insulator without a spectral gap.

We thank G. Vignale for illuminating discussions. Work partially supported by the ONR Grant No. N00014-12-1-1041. 
Computer facilities were provided by the project Equip@Meso (reference ANR-10-EQPX-29- 01).

\end{document}